\documentclass[pdftex,twocolumn,9pt,letterpaper]{extarticle}

\newcommand{\AUTHORS}{Haoyu Zhang, Logan Stafman, Andrew Or, Michael J. Freedman}
\newcommand{\TITLE}{SLAQ: Quality-Driven Scheduling for Distributed Machine Learning}
\newcommand{\KEYWORDS}{Scheduling, Machine Learning, Approximate Computing, Resource Management, Quality}
\newcommand{\CONFERENCE}{In the 1st SysML Conference, Stanford, CA, February 2018}
\newcommand{\CATEGORIES}{[\emph{Computer systems organization}]: Distributed architectures; [\emph{Computing methodologies}]: Distributed artificial intelligence; [\emph{Theory of computation}]: Approximation algorithms analysis}
\newcommand{\PAGENUMBERS}{yes}       
\newcommand{\COLOR}{yes}
\newcommand{\showComments}{yes}
\newcommand{\comment}[1]{}
\newcommand{\onlyAbstract}{no}
\newcommand{\name}{\small \sf SLAQ}

\usepackage[T1]{fontenc}
\usepackage[utf8]{inputenc}
\usepackage{bm}                        
\usepackage[scaled=0.92]{helvet}
\usepackage{courier}
\usepackage{amsmath}
\usepackage{amsthm}
\usepackage{times}
\usepackage{mathptmx}

\usepackage{ifthen}
\ifthenelse{\equal{\PAGENUMBERS}{yes}}{%
\usepackage[nohead,
            left=1in,right=1in,top=1in,
            footskip=0.5in,bottom=1in,     
            columnsep=0.25in
            ]{geometry}
}{%
\usepackage[noheadfoot,left=1in,right=1in,top=1in,
            footskip=0.5in,bottom=1in,
            columnsep=0.25in
      ]{geometry}
}

\usepackage[font={footnotesize}]{caption}
\usepackage[small,compact]{titlesec}

\titlespacing\subsection{0pt}{10pt plus 4pt minus 2pt}{8pt plus 2pt minus 2pt}
\titlespacing\subsubsection{0pt}{8pt plus 4pt minus 2pt}{6pt plus 2pt minus 2pt}

\makeatletter
\renewcommand{\paragraph}{%
  \@startsection{paragraph}{4}%
  {\z@}{0.45ex \@plus .2ex \@minus .1ex}{-1em}%
  {\normalfont\normalsize\bfseries}%
}
\makeatother

\usepackage{enumitem}
\setlist{itemsep=0pt,parsep=0pt}             

\usepackage{algorithm}
\usepackage[noend]{algpseudocode}

\usepackage{mathtools}

\usepackage{subfigure}

\usepackage{fancyhdr}

\ifthenelse{\equal{\PAGENUMBERS}{yes}}{%
  \pagestyle{plain}
}{%
  \pagestyle{empty}
}

\usepackage[numbers]{natbib}

\usepackage{booktabs}
\usepackage{color}
\usepackage{colortbl}
\usepackage{float}                           

\ifthenelse{\equal{\COLOR}{yes}}{%
  \usepackage[colorlinks,citecolor=blue]{hyperref}
}{%
  \usepackage[pdfborder={0 0 0}]{hyperref}
}
\usepackage{url}

\hypersetup{%
pdfauthor = {\AUTHORS},
pdftitle = {\TITLE},
pdfsubject = {\CATEGORIES. \CONFERENCE},
pdfkeywords = {\KEYWORDS},
bookmarksopen = {true}
}

\definecolor{placeholderbg}{rgb}{0.85,0.85,0.85}

\usepackage[pdftex]{graphicx}
\usepackage{soul}
\soulregister\cite7
\soulregister\ref7
\soulregister\pageref7

\usepackage{flushend}

\usepackage[absolute]{textpos}

\usepackage{textcomp}
\floatstyle{plain}
\newfloat{acmcr}{b}{acmcr}

\newcommand{\note}[2]{
    \ifthenelse{\equal{\showComments}{yes}}{\textcolor{#1}{#2}}{}
}

\date{}
\title{{\bf \fontsize{15}{18}\selectfont \TITLE\footnote{This work has been
previously published in ACM SoCC '17~\cite{slaq-socc17}.}}}
\author{{Haoyu Zhang, Logan Stafman, Andrew Or, Michael J. Freedman}\\
        {\em Princeton University} }
\begin{document}

\maketitle

\ifthenelse{\equal{\PAGENUMBERS}{yes}}{%
  \thispagestyle{fancy}
}{%
  \thispagestyle{empty}
}


\ifthenelse{\equal{\onlyAbstract}{no}}{%
\section{Background and Motivation}
\label{sec:background}


Machine learning (ML) is an increasingly important tool for 
large-scale data analytics.
A key challenge in analyzing massive amounts of data with ML arises from the
fact that model complexity and data volume are growing much faster than hardware speed improvements.
Thus, time-sensitive ML on large datasets necessitates the
use and efficient management of cluster resources.
Three key features of ML are particularly relevant to resource management.


\paragraph{ML algorithms are intrinsically approximate.}
ML models are approximate functions for input-output mapping.
We use \emph{quality} to measure how well the model maps input to the correct output.
Training ML models is a process of optimizing the model parameters to
maximize the quality on a dataset.

\paragraph{ML training is typically iterative with diminishing returns.}
Algorithms such as Gradient Descent, L-BFGS and
Expectation Maximization (EM) are widely used to \emph{iteratively}
solve the numerical optimization problem.
The quality improvement \emph{diminishes} as more iterations are completed (Figure~\ref{fig:norm-time}).

\paragraph{Training ML is an exploratory process.}
ML practitioners retrain their models repeatedly to explore feature validity~\cite{feature-engineering-cidr13},
tune hyperparameters~\cite{practical-bayesian-nips12, gradient-based-icml15, hyperband-arxiv16, population-arxiv17},
and adjust model structures~\cite{deep-compression-corr15}, in order to operationalize the final model with the best quality.
Practitioners in experimental
environments often prefer to work with more approximate models (e.g., 95\% loss reduction) trained within a short
period of time for preliminary testing, rather than wait a
significant amount of time for a perfectly converged model with poorly tuned
configurations.  


Existing schedulers primarily focus on \emph{resource fairness}~\cite{yarn,
mesos-nsdi11, drf-nsdi11, hierarchical-socc13, hadoop-capacity, quincy-sosp09},
but are agnostic to model quality and resource efficiency.
With this policy, equal resources will be allocated to jobs
that are in their early stages and could benefit significantly from extra
resources as those that have nearly converged and cannot improve much further.
This is not efficient.
The key intuition behind our system is that \emph{in the context of approximate ML
training, more resources should be allocated to jobs that have the most
potential for quality improvement}.

\section{Design}
\label{sec:design}

\begin{figure*}[t!]
\begin{minipage}[t][][b]{0.3\textwidth}
  \includegraphics[width=\textwidth]{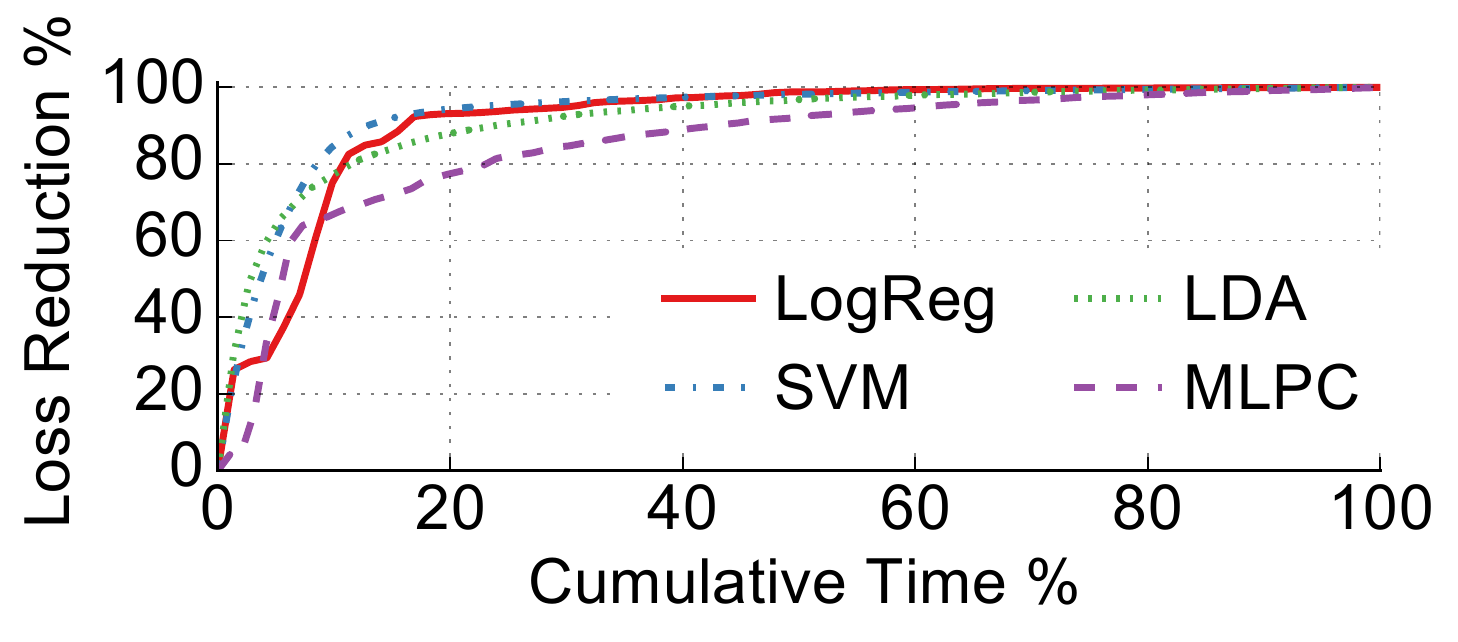}
  \caption{$>80\%$ of work is done in $<20\%$ of time.}
  \label{fig:norm-time}
\end{minipage}
\hspace{.2ex}
\begin{minipage}[t][][b]{0.32\textwidth}
  \includegraphics[width=\textwidth]{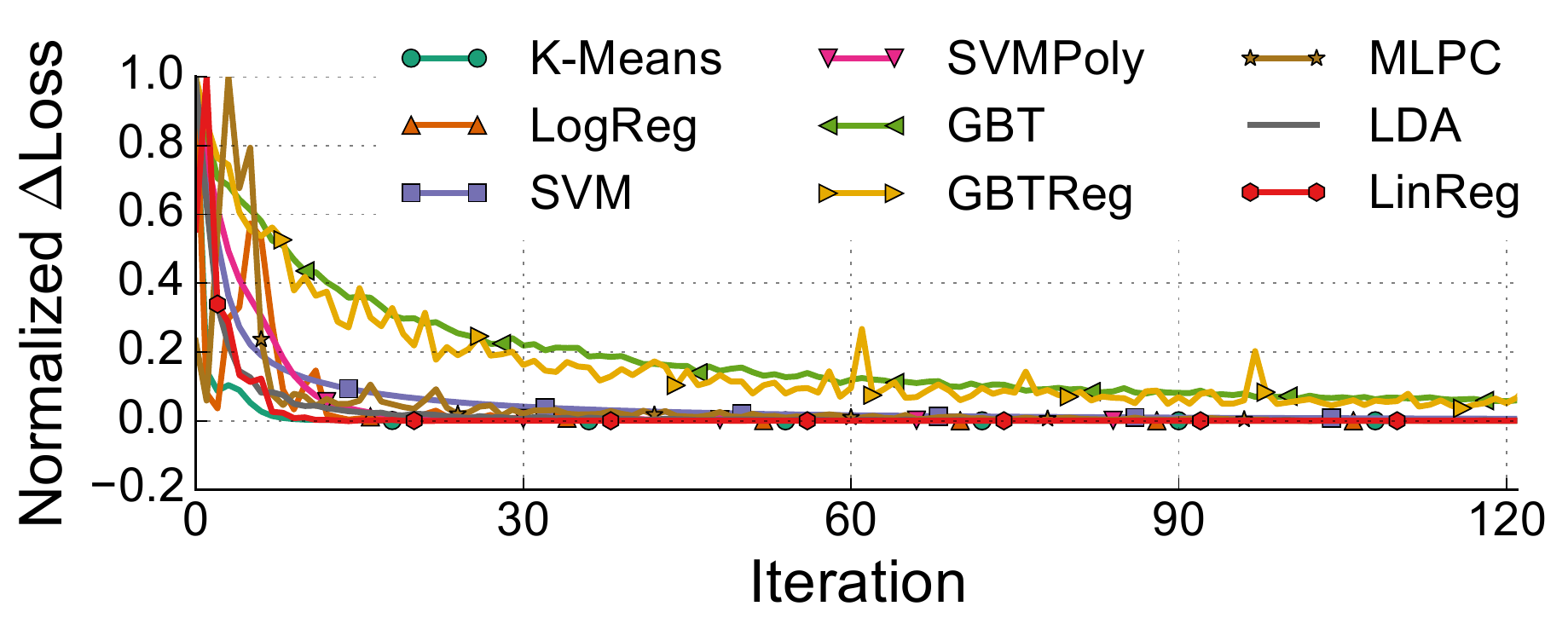}
  \caption{Normalized $\Delta$Loss for ML algorithms.}
  \label{fig:norm-loss}
\end{minipage}
\hspace{.2ex}
\begin{minipage}[t][][b]{0.36\textwidth}
  \includegraphics[width=\textwidth]{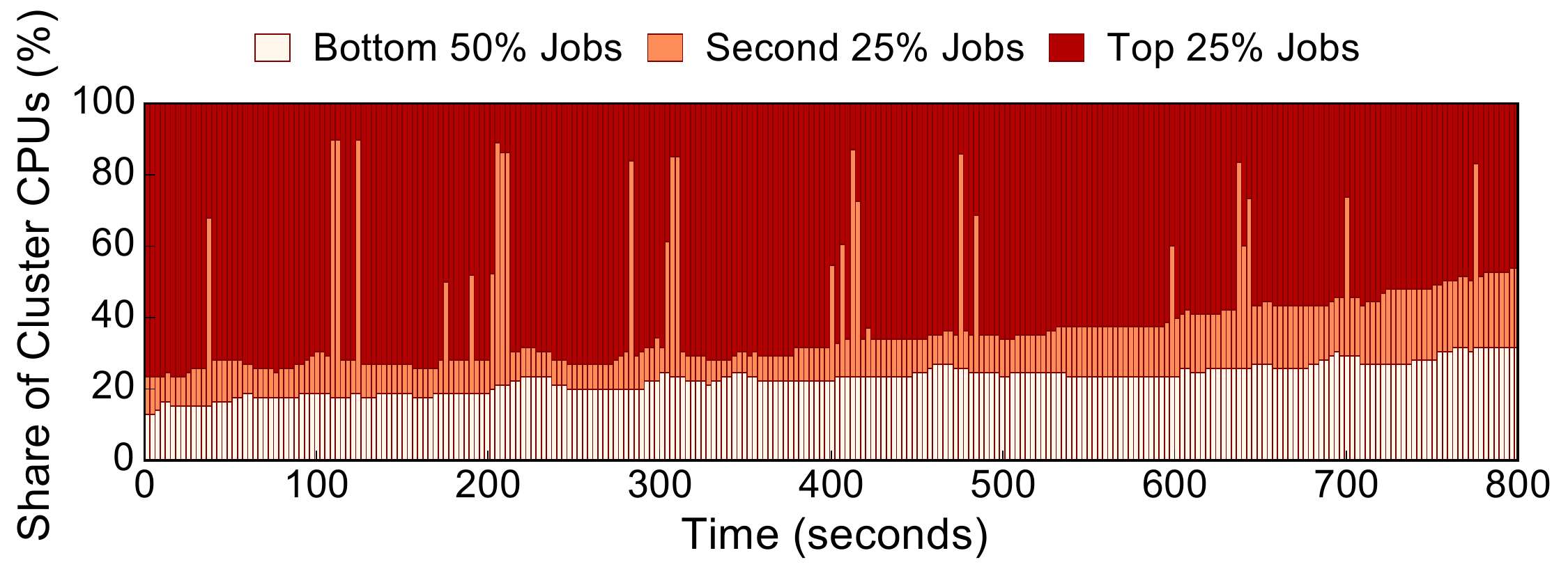}
  \caption{Resource allocation across job groups.}
  \label{fig:cpu-dist}
\end{minipage}
\end{figure*}

\begin{figure*}[t!]
\begin{minipage}[t][][b]{0.3\textwidth}
  \centering
  \includegraphics[width=\textwidth]{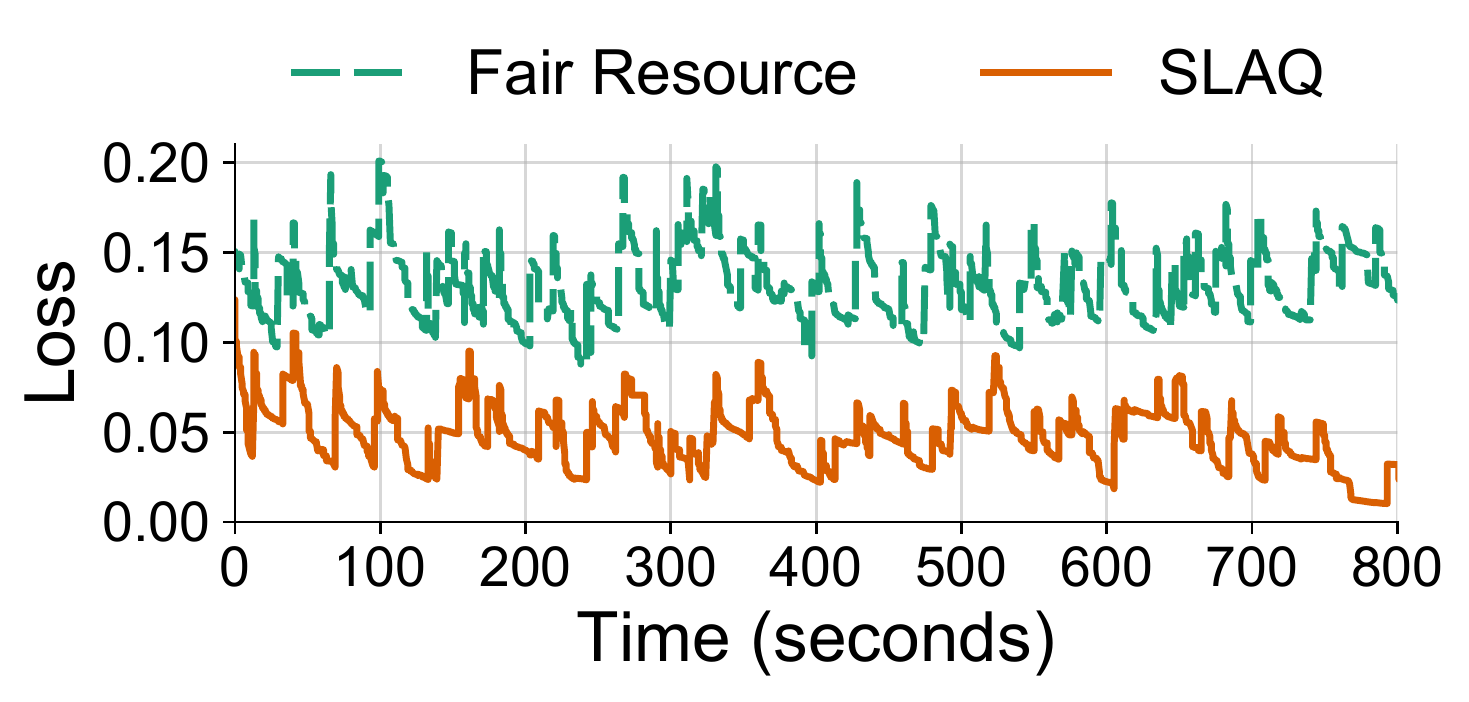}
  \caption{Average of normalized loss values.}
  \label{fig:loss-improve1}
\end{minipage}
\hspace{2ex}
\begin{minipage}[t][][b]{0.33\textwidth}
    \centering
  \includegraphics[width=.9\textwidth]{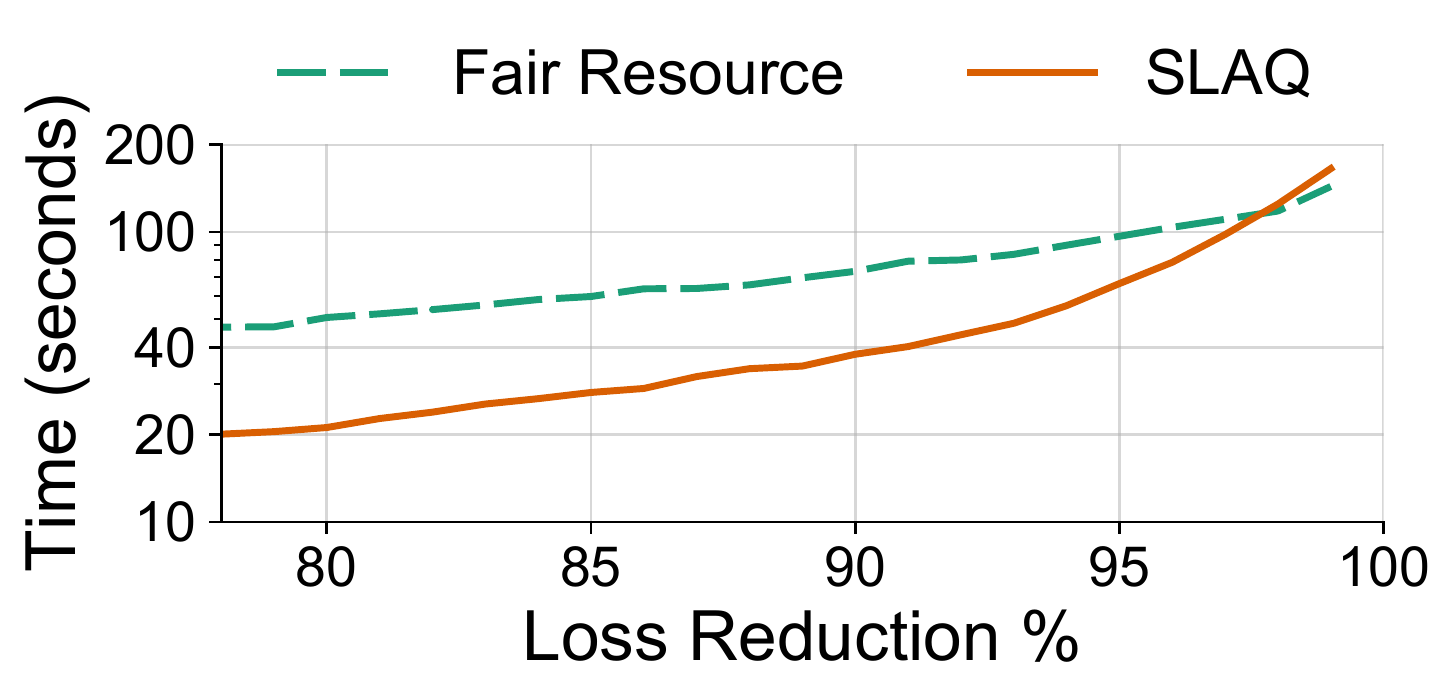}
  \caption{Time to achieve loss reduction percentage.}
  \label{fig:loss-improve2}
\end{minipage}
\hspace{2ex}
\begin{minipage}[t][][b]{0.3\textwidth}
  \includegraphics[width=.92\textwidth]{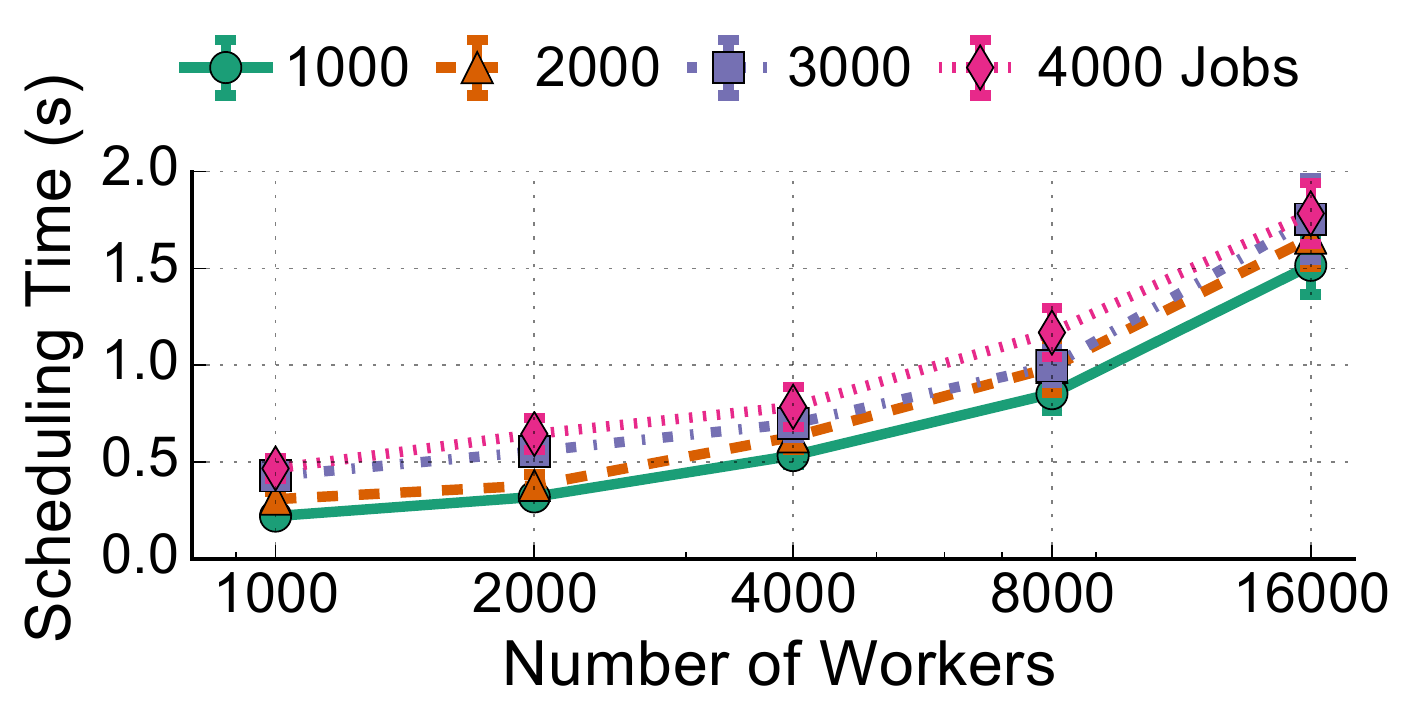}
  \caption{Scheduling time.}
  \label{fig:scalability}
\end{minipage}
\end{figure*}

We present {\name}, a cluster scheduling system for ML training jobs that aims
to maximize the overall job quality.
To achieve this, {\name} needs to
(1) normalize the quality metrics in order to trades off resources and quality
across multiple jobs;
(2) predict how much progress the job would achieve if it was granted a certain amount of resources;
(3) efficiently allocate cluster CPUs to maximize the system-wide quality
improvement.

\paragraph{Normalizing Quality Metrics.}
While metrics like accuracy and F1 score~\cite{f1score} are intuitively
understandable, they are not applicable to non-classification algorithms.
In contrast, loss functions are internally calculated by almost all 
algorithms in each iteration, but each loss function has a different
real-world interpretation, and its range, convexity, and monotonicity of
depend on both the models and the optimization
algorithms. Directly normalizing loss values
requires a priori knowledge of the loss range, which is impractical in an online setting.

We choose to normalize the \emph{change} in loss values between iterations, with respect
to the largest change we have seen so far.
Figure~\ref{fig:norm-loss} shows the normalized changes of loss values for
common ML algorithms.  Even though the algorithms have diverse loss
ranges, we observe that the changes generally follow similar convergence properties,
and can be normalized to decrease from $1$ to $0$. This helps {\name} track and
compare the
progress of different jobs, and, for each job, correctly project the
time to reach a certain loss reduction with a given resource allocation.
Note that this approach currently does \emph{not} support some non-convex algorithms (such as training Deep Neural Networks) due to the
lack of convergence of these analytical models.

\paragraph{Predicting Quality Improvement.}
Previous work~\cite{ernest-nsdi16, cherrypick-nsdi17} estimates general-purpose
big-data job runtime by analyzing the job computation and
communication structure, using offline analysis or code profiling.
As the computation and communication pattern changes during ML model configuration tuning, the process
of offline analysis needs to be performed every time, thus incurring significant
overhead.

We use \emph{online} quality prediction by leveraging the convergence properties
of the loss functions. Based on the optimizers used for minimizing the loss
function, we can broadly categorize the algorithms by their convergence
rate.

I. \emph{Algorithms with sublinear convergence rate.}
First-order algorithms\footnote{Assume $f$ is convex, differentiable, and $\nabla f$ is Lipschitz continuous.}
(e.g., gradient descent)
have a convergence rate of $O(1/k)$, 
where $k$ is the number of iterations~\cite{statistical-learning}.
The convergence rate could be
improved to $O(1/k^2)$ with optimization.

II. \emph{Algorithms with linear or superlinear convergence rates.}
Algorithms in this category\footnote{Assume $f$ is convex and twice continuously differentiable, optimizers can take advantage of the second-order derivative to get faster convergence.}
have a convergence rate of $O(\mu^k), |\mu| < 1$.
For example, L-BFGS, which is a widely used Quasi-Newton
Method, has a superlinear convergence rate which is between linear and
quadratic. 

With the assumptions of loss convergence rate, 
we use exponentially weighted history loss values at
to fit a curve
$f(k) = \frac{1}{ak^2 + bk + c} + d$ for sublinear algorithms, or $f(k) = \mu^{k-b} + c$ for
linear and superlinear algorithms.
Intuitively, loss values obtained in the near past are more informative
for predicting the loss values in the near future.
Experiments show that this prediction achieves less than $5\%$ prediction errors for all the algorithms in Figure~\ref{fig:norm-loss} when predicting the next 10th iteration.


\paragraph{Scheduling Based on Quality Improvements.}

We schedule a set of $J$ jobs running concurrently on the shared cluster for a
fixed scheduling epoch $T$. The optimization problem for maximizing the total normalized loss reduction
over a short time horizon $T$ is as follows. Sum of allocated resources $a_j$
cannot exceed the cluster resource capacity $C$.
\vspace{-5pt}
\begin{eqnarray}
  \max_{j \in J} & \sum_j Loss_j(a_j, t) - Loss_j(a_j, t + T) \nonumber \\
  s.t.           & \sum_j a_j \leq C                          \nonumber
\end{eqnarray}


\vspace{-5pt}
The algorithm starts with $a_j = 1$ for each job to prevent starvation.
At each step we consider increasing $a_i$ (for all jobs $i$) by one unit
(i.e., one CPU core) and get the predicted loss reduction. Among these jobs,
we pick the job $j$ that gives the highest loss reduction, and increase $a_j$ by
one unit. We repeat this until we run out of available resources.

\section{Evaluation}
\label{sec:eval}


\paragraph{Setup.}
We implemented
{\name} within the Apache Spark framework~\cite{spark-nsdi12} and utilize its accompanying
MLlib machine learning library~\cite{mllib-corr15}. 
Our testbed consists of a cluster of 20 {\sf c3.8xlarge} EC2 instances
on the AWS Cloud.
We tested {\name} with the most common ML algorithms, including
(i) classification: SVM, Neural Network (MLPC), Logistic Regression, GBT, and our extension to Spark MLlib with SVM polynomial kernels;
(ii) regression: Linear/GBT Regression;
(iii) unsupervised learning: K-Means, LDA.
Each algorithm is further diversified to construct different models.
We collected more than $200$GB datasets
from various online sources, spanning numerical, plain texts~\cite{associated-press},
images~\cite{mnist}, audio meta features~\cite{million-song}, and so on~\cite{libsvm}.
The baseline we compare against is a work-conserving fair scheduler, which is the
widely-used scheduling policy for cluster computing frameworks~\cite{yarn, mesos-nsdi11, drf-nsdi11, hadoop-capacity, quincy-sosp09}.

\paragraph{Scheduler Quality and Runtime Improvement.}
We submit a set of $160$ ML training jobs
with different models, following a Poisson distribution (mean arrival time $15$s).
Figure~\ref{fig:loss-improve1} shows the average normalized loss values across
running jobs in an $800$s time window of the experiment.
When a new job arrives, its initial loss is $1.0$, raising the average loss value;
the spikes indicate new job arrivals. 
The average loss value achieved by {\name} is on average $73\%$ lower than that of the
fair scheduler.

Figure~\ref{fig:loss-improve2} shows the average time it takes a job to achieve
different loss values. As {\name} allocates more resources to jobs that have
the most potential for quality improvement, it reduces the average time to reach $90\%$
($95\%$) loss reduction from $71$s ($98$s) down to $39$s ($68$s), $45\%$ ($30\%$)
faster.
For exploratory training, this level of accuracy is frequently sufficient.
Thus, in an environment where users submit exploratory ML training jobs, {\name}
could substantially reduce users' wait times.

Figure~\ref{fig:cpu-dist} explains {\name}'s benefits by plotting the allocation of
CPU cores in the cluster over time. Here we group the active jobs by their normalized loss:
(i) $25\%$ jobs with high loss values;
(ii) $25\%$ jobs with medium loss values;
(iii) $50\%$ jobs with low loss values (almost converged).
With a fair scheduler, the cluster CPUs should be allocated to the three groups
proportionally to the number of jobs.
In contrast, {\name} allocates much
more resource to (i) ($60\%$) than to (iii) ($22\%$), which is the
underlying reason for the improvement in Figures~\ref{fig:loss-improve1}
and \ref{fig:loss-improve2}.

\paragraph{Scalability and efficiency.}
{\name} is a \emph{fine-grained} job-level scheduler: it 
allocates resources between competing ML \emph{jobs},
but does so over \emph{short time intervals} to ensure the continued rebalancing of
resources across jobs.
Figure~\ref{fig:scalability} plots the time to
schedule tens of thousands of concurrent jobs on large clusters
(simulating both the jobs and worker nodes).  {\name} makes its
scheduling decisions in hundreds of milliseconds to a few seconds,
even when scheduling $4,000$ jobs across 16K worker cores.
{\name} is sufficiently fast and scalable for (rather aggressive) real-world
needs.

\section{Conclusion and Future Work}
\label{sec:concl}

{\name} is a quality-driven scheduling system designed for
large-scale ML training jobs in shared clusters.
{\name} leverages the iterative nature of ML algorithms and obtains
highly-tailored prediction to maximize the quality of models produced by a
large class of ML training jobs.
As a result, {\name} improves the overall quality of
executing ML jobs faster, particularly under resource contention.

\paragraph{Non-convex optimization.}
Loss functions of non-convex optimization are not guaranteed to
converge to global minima, nor do they necessarily decrease monotonically.
The lack of an analytical model of the convergence properties
interferes with our prediction mechanism, causing {\name} to underestimate
or overestimate the potential loss reduction.
One potential solution is to let users provide the scheduler with
hint of their target loss or performance, which could be acquired from
state-of-the-art results on similar problems or previous training trials.
The convergence properties of non-convex algorithms is being actively
studied in the ML research community~\cite{nonconvex1, nonconvex2}.
We leave modeling the convergence of these algorithms to future work,
and an interesting topic for future discussion at SysML.



\setlength{\bibsep}{2pt plus 1pt}  
\small 
\bibliography{ref}
\bibliographystyle{abbrvnat_noaddr} 
}{
}

\end{document}